\documentclass[10pt]{article}
\pdfoutput=1
\usepackage{amsfonts}
\usepackage[english]{babel}
\usepackage[para]{threeparttable}
\usepackage{graphicx}
\usepackage{pdfpages}
\usepackage{nccmath}
\usepackage{bmpsize}
\usepackage{caption}
\usepackage{geometry}
\newcommand{\BEQ}{\begin{equation}}
\newcommand{\EEQ}{\end{equation}}
\newcommand{\NEQ}{\end{equation}}

\renewcommand{\theequation}{\thesection.\arabic{equation}}
\makeatletter
\@addtoreset{equation}{section}
\makeatother
\begin{document}
\bibliographystyle{plain}
\title{
\bf  Combinatorial Models of Cross-Country Dual Meets:  What is a Big Victory?
}

\author{
Kurt S.~Riedel
\thanks{Millennium Partners,
666 Fifth Ave, New York, NY 10103}
} %\end{center}

\date{June 28, 2019}
%\today
\maketitle

\begin{abstract}
Combinatorial/probabilistic models for cross-country dual-meets are proposed. The first model assumes
that all runners are equally likely to finish in any possible order. The second model assumes that each
team is selected from a large identically distributed population of potential runners and with each potential runner's ranking determined by the initial draw from the combined population.   
\end{abstract}

68R05 Combinatorics 97K20 Combinatorics 97K80 Applied statistics 97M40 Operations research, economics 05A15 Exact enumeration problems, generating functions

%\newpage

\section{Introduction} \label{I}
%\graphicspath{{/home/kriedel/Tex/CCFig/}}
%\graphicspath{{/home/kriedel/Tex/CCfinal/}}
Cross-country racing is a varsity sport in many high schools with more than 15 thousand schools offering cross-country \cite{NFHS}. It is the fourth most popular sport for boys and fifth most for girls, as defined by the number of school teams. 
Almost half a million high school students compete on cross-country teams every year, and hundreds of thousands of middle school students also compete.
Scoring is an application
of nonparametric statistics: all that matters is the rank of the runner as he/she crosses the finish line.
The team score is the sum of the ranks of the runners whose score counts. This scoring is termed ``Rank Sum Scoring.''
In this article, we calculate the distribution of scores under a variety of assumptions on the
relative speed of the runners. First, we address the distribution of scores when all runners have the same distribution of 
running times.

To the best of our knowledge, the distribution of outcomes has never been
analyzed from probabilistic point of view. There is much excellent research on cross-country
scoring from a game theoretical perspective \cite{Huntington}, \cite{Hammond}, \cite{Sze},\cite{MixonKing},
\cite{BESW2014}, \cite{BS}, \cite{BERS}. The basic discrepancy/inequity is that in a multi-team meet,
Team A can beat Team B under dual-meet scoring and lose to it under the multi-team scoring rule.
These papers analyze the multi-team scoring and propose alternative scoring rules that eliminate these paradoxes.
In \cite{Hammond} and \cite{BESW2014}, distributions of scores are calculated for meets involving three or more teams
and the probability of a ``social choice principle violation'' is calculated.  
In \cite{BERS}, actual multi-team empirical data is analyzed to see how often these scoring
paradoxes occur in practice. They find that combinatorial analyses are helpful in predicting the probability of a paradox.
Their results lend credence to our combinatorial analysis.

%We know of only one previous mathematical analysis of cross-country.
%In \cite{Sze}, Szydlik uses game theory to analyze alternative scoring rules to try to find an equilibrium scoring system where the results of multi-team invitational meets are consistent with the results of equivalent dual-meets between the various teams.
%The results in \cite{Sze} are interesting from a game theory perspective,
The standard scoring method is well established and unlikely to change. dual-meet scoring is based purely upon
the order that the runners finish the race. Our study examines the standard rank-based scoring in a cross-country dual-meet. We derive the actual scoring distributions for the standard scoring rules for dual-meets
using combinatorics and rank statistics. Indeed, we believe that the nonparametric (rank-based) nature of cross-country scoring makes it an ideal example for teaching stochastic and combinatorial analysis.

%Our methods are part combinatorial and part 
%computational. None of our results are difficult to obtain. There are many generalizations and simplifications. We
%believe that cross-country scoring analyses along the lines of this article constitute 
%a set of fun and educational problems for undergraduate courses and possibly high school classes.

Definition: An $(M,N)$ dual meet consists of a team of $M$ runners. Only the $N$ fastest runners on each team are counted. The meet is scored as follows: the $k$th runner to finish receives a a score of $k$. The placement of runners from $N+1$ to $M$ are not included in the score but do increase the score of the other team. The $N$ lowest scores are added up for each team. The lowest score wins. If the runners $N+1$ to $M$ do not raise the score of the
other team, we refer to the event as a cross-country meet with {\it no displacement}.

Cross-country meets occur with and without displacement scoring, but displacement scoring is definitely more common. Thus  
we make displacement scoring the default and explicitly mention {\it no displacement} in the cases where we study this alternative.
For an $(M,N)$ event, the sum of the two scores is between $N*(2N+1)$ and $N*(M+N+1)$.
The upper bound score, $N*(M+N+1)$, occurs when all $M$ runners of one team are faster than any of the other team.
For the  no displacement case, the sum of the two scores is $N*(2N+1)$.

The standard American dual meet is a $(7,5)$ with seven runners and lowest five scores are counted. An international dual meet has six runners and four are counted.  We will consider two teams, Team A and Team B. In certain cases,
we will assume that Team A has the fastest runner or even the fastest two runners. We will denote the score of Team A by $s_A$ and the score
of Team B by $s_B$. The margin of victory for Team A will be denoted by $m_{A,B}= s_B -s_A$.

The following results are well known to the cross-country community.

1) The best possible score is $s_A= N*(N+1)/2$. The worst possible score is $s_A= MN + N*(N+1)/2$

2) The biggest possible victory margin is $MN$. For $M=N=4$, this is 16, and for $M=N=5$, the maximum victory margin is $25$.
For $(M,N)=(7,5)$, the maximum victory margin is $35$. For $(M,N)=(6,4)$, the maximum victory margin is $24$.

3) If $N=5$ and $M \le 7$ and Team A has the fastest three runners, Team A has won.
Proof: $(1+ 2+ 3 +11+12)< (4+5+6+7+8)$.

For no displacement scoring, we have the following results

A) The sum of the two teams' scores satisfies $s_A +s_B= N *(2N+1)$. For $N=4$ and $N=5$, $s_A +s_B= 36$ and $55$ respectively.

B) Ties can occur only if $N$ is even. In particular, ties can occur for $(6,4)$, but not for $(7,5)$.

C) The score difference is always odd for $(7,5)$. The score difference for $(6,4)$ is always even.

D) If $M=6$, $N=4$ and Team A has the fastest two runners, Team A cannot lose. Proof: $1+2+7+8=3+4+5+6$
%1+2+9+10> 3+4+5+6
%Results 1) and 2) follow from summing $1$ through $N$ and $N+1$ to $2N$. Result 3) is the sum of $1$ through $2N$.
%Results 4) and 5) follow from $s_B-s_A =  N *(2N+1) - 2*s_A$. Result 3 follows from $ 1+2+7+8=3+4+5+6$. Result 4
%follows from $1+2+3+9+10<4+5+6+7+8$.

As pointed out by the anonymous referee, the no displacement case is a reparameterization of Wilcoxon rank-sum distribution
(Mann-Whitney distribution) \cite{Wilcoxon}, \cite{MannWhitney}, \cite{Lehman}, \cite{Conover}. We calculate the distribution of
$s_B-s_A = 2 s_B - N *(2N+1)$. The case with displacement seems to be a related but new distribution.
In the next sections, we calculate the distribution of victory margins, both one-sided and two sided.
We define a large victory to be a score difference (positive or negative) in the .9 quantile range or greater.
This means ten percent of the dual-meets will result in a ``large'' victory by our definition.

\section{I.I.D. Outcome Distribution}
%\graphicspath{{/home/kriedel/Tex/CCFig/},{/home/kriedel/Tex/},{/home/kriedel/Tex/CCver1/}}

We begin by considering the case where all runners are equally likely to come in $k$th place in the competition for $k$ in $1, \ldots ,2M$. In other words, 
the runners' times are independently, identically distributed (IID). We label the places from $1$ to $2M$. We then draw $M$ of these without replacement
to determine the the placement of the runners on Team A.
There are ${2M \choose M}$ possible outcomes.

To illustrate the distribution and scoring, we consider the three runner case with two scores counting i.e. $(3,2)$.
% possible ranks of a three person team dual-meet ;X=crossC.fillCases(3); Y=crossC.dropLast(X,2); X.tolist();map(tuple, X.tolist())
The list of possible outcomes for Team A is
$[(1, 2, 3), (1, 2, 4), (1, 2, 5)$, $(1, 2, 6), (1, 3, 4)$, $(1, 3, 5), (1, 3, 6), (1, 4, 5)$, $(1, 4, 6)$, $(1, 5, 6)$, $(2, 3, 4)$, $(2, 3, 5)$, $(2, 3, 6), (2, 4, 5)$, $(2, 4, 6), (2, 5, 6), (3, 4, 5)$, $(3, 4, 6), (3, 5, 6), (4, 5, 6)]$.
By assumption, all cases are equally likely.
Since only the top two runners count, the corresponding list of scores is
distribution of scores for top 2 of each team.
When there is no displacement, the distribution of scores is
$\{(1, 2):4, (1, 3):3, (1, 4):3, (2, 3):3, (2, 4):3, (3, 4):4 \}$.
Thus probability of a tie is $0.3$, the probability of a two point victory is $0.3$ and the probability of a four point victory is $0.4$.
The team with the fastest runner cannot lose and ends up winning $70$ percent of the time.
Now consider the case with displacement. We present the scores as a tuple, $(s_A,s_B)$ for each scenario,
$[(3,9),(3, 8), (3,7), (3,7), (4,7), (4,6), (4,6), (5,5), (5,5)$, $(6,5),(5,6),(5,5),(5,5)$, $(6,4),(6,4),(7,4),(7,3),(7,3),(8,3),(9,3)]$.
The distribution of score differences, $s_B-s_A$ is symmetric and satisfies 
$\{0:4,1:1,2:2,3:1, 4:2, 5:1,6:1\}$. Not surprisingly, displacement increases the diversity of outcomes and can have larger victories.

Let us consider the alternative case of two runners on each team and both runners count, $(2,2)$. 
Now, all six cases are equally likely: $[(1, 2), (1, 3), (1, 4), (2, 3), (2, 4), (3, 4)]$.
Here the probability of a tie is $1/3$, the probability of winning by two points is $1/3$ and the probability of a four point victory is $1/3$.
Finally, consider the case of $(M,2)$ for large $M$. As $M$ increases, the distribution of scores converges to the distribution
corresponding to drawing with replacement: the probability that one team has the two fastest runners is $0.5$. If there is no displacent, the probability of a tie is $0.25$ and the probability of a two point margin is $0.25$. With displacement, let Team A have the fastest runner. Team A has a score distribution:$\{ 3:.5,4:.25,5:.125,6:.0625\ldots \}$
for a mean score of 4. Team B has a score distribution:$\{5:.25,6:.125,7:.1875 \ldots \}$ for an expected score of $8$. (The first Team B runner has an expected score of three while the second Team B runner has an expected score of five.)
This illustrates a common theme: the larger $M$ for a given $N$, the larger the tail event scores are. % involving large victory margins.

These three cases illustrate how the distribution of scores vary as the assumption of number of runners varies.
We now focus on the two real-world situations: international dual-meets, $(6,4)$ and American dual-meets, $(7,5)$.
Rather than trying to derive the score distribution analytically, we write a very short program which evaluates the score
for each of the ${2M \choose M}$ possible outcomes. In Python, the outcomes are given by $map(tuple,itertools.combinations(range(2*M), M))$,

%In Table \ref{tabRun4},
We tabulate the score difference, $s_B-s_A$, conditional on Team A having the fastest runner.
Remember the team with the lowest score wins, so  $s_B-s_A$  is the victory margin for Team A. A negative score means that Team A lost. % despite having the fastest runner.
We begin by computing the distribution with no score displacement in Table \ref{tabRun4}.
There are only 13 distinct values of score difference conditioned on Team A having the fast runner and the maximum victory margin being 16.
%Table \ref{tabRun4} has two rows of score differences to accomodate all the possible outcomes.
%In Table \ref{tabRun4}, we assume that Team A has the fastest runner.
Given the fastest runner, Team A  wins with probability $0.6948$ and ties with probability $0.0974$.
Conditional on Team A winning and having the fastest runner, the mean victory margin is $8.11215$.
Team A loses with probability $0.20779$ and the average loss is $4.45833$.
\begin{table}[!ht] %[ht]
\caption{Distribution of score differences Runner:6 Score:4  No Displacement}
% title of Table
\centering 
\begin{tabular}{cccccccc}
\hline \hline 
$s_B-s_A$ &-8&-6&-4&-2& 0& 2& 4\\
counts &21&15&25&35&45&40&55\\
prob &.0455&.0325&.0541&.0758&.0974&.0866&.1190\\
\hline
$s_B-s_A$ & 6& 8&10&12&14&16& \\
counts &45&50&46&36&21&28 & \\
prob &.0974&.1082&.0996& .0779&.0455&.0606 &\\
\hline \hline
\end{tabular}
\label{tabRun4}
\caption*{Distribution conditional on Team A having the fastest runner without displacement.}
%\begin{tablenotes}
%\item 
%Distribution conditional on Team A having the fastest runner
%\end{tablenotes}
\end{table}
To get the unconditional distribution, $|s_A-s_B|$ (either Team A or Team B may have the fastest runner),
we just symmetrize Table \ref{tabRun4}. The symmetrized distribution is a reparameterization of the Wilcoxon rank-sum distribution.
Unconditionally, the mean score difference is $6.56277 \pm 4.52355$ and median victory is $6$. The $75$th percentile victory is $10$.
A reasonable definition of a large victory is the $90\%$ quantile of the score difference. We see that this quantile value is  $14$. 

Table \ref{tabRun4D} has the more realistic case of displacement. There are now 39 possible values of the score difference.
The probability of an odd score difference is nonzero but clearly smaller than the probability of an even score difference.
\begin{table}[!ht] %[ht]
\caption{Distribution of score differences Runner:6 Score:4 with Displacement}
% title of Table
\centering 
\begin{tabular}{ccccccccccc}
\hline \hline 
$s_B-s_A$&-14&-13&-12&-11&-10& -9& -8& -7& -6& -5\\
counts & 1& 1& 2& 4& 5& 4&11& 4&12& 7\\
prob & .0022&.0022&.0043&.0087&.0108&.0087&.0238&.0087&.0260&.0152\\
\hline
$s_B-s_A$ & -4& -3 &-2& -1&  0&  1&  2&  3&  4&  5\\
counts &18& 7&24&12&27& 9&27&12&33&12\\
prob & .0390&.0152&.0519&.0260&.0584&.0195&.0584&.0260&.0714&.0260\\
\hline
$s_B-s_A$ &  6&  7&  8&  9& 10& 11& 12& 13& 14& 15\\
counts &31&13&31&10&28&11&23&11&17& 9\\
prob & .0671&.0281&.0671&.0216&.0606&.0238&.0498&.0238&.0368&.0195\\
\hline
$s_B-s_A$ &  16& 17& 18& 19& 20& 21 &22& 23& 24 \\
counts &14& 8& 7& 5& 6& 2& 2& 1& 1 \\
prob & .0303&.0173&.0152&.0108&.0130&.0043&.0043&.0022&.0022 \\
\hline
\hline \hline
\end{tabular}
\label{tabRun4D}
\caption*{Distribution conditional on Team A having the fastest runner with displacement.}
\end{table}
Table \ref{tabRun4D} has four rows of score differences to accomodate all the possible outcomes. In Table \ref{tabRun4D}, we assume that Team A has the fastest runner. Given the fastest runner, Team A  wins with probability $0.6991$ and ties with probability $0.0584$.
Conditional on Team A winning and having the fastest runner, the mean victory margin is $9.08$.
Team A loses with probability $0.30$ and the average loss is $4.973$. Thus, displacement makes the importance of the fastest runner in determining
which team wins significantly less, because the slower runners have more influence.

To get the unconditional distribution (either Team A or Team B may have the fastest runner) for $(6,4)$ with displacement, we just symmetrize Table \ref{tabRun4D} to get the distribution of $|s_A-s_B|$.
Unconditionally, the mean score difference is $7.554 \pm 5.334$ and median score difference is $6.5$. The $75$th percentile of the score difference is $11$.
A reasonable definition of a large victory is the $90\%$ quantile of the absolute score difference. We see that this quantile value is  $15$.

As a variant of the basic distribution, let us consider the scoring distribution of $(6,4)$ conditional on Team A having the fastest two runners.
This result is given in Table \ref{tabRun4b} for the case of no displacement.
In this case, the probability of a tie is $0.0714$.
The mean score differential is $9.162\pm 4.711$ and the median difference is $10$.
\begin{table}[ht]
\caption{Distribution of score differences Runner:6 Score:4 (No Displacement)}
% title of Table
\centering 
%\hline
\begin{tabular}{cccccccccc}
\hline 
$s_B-s_A$& 0& 2& 4& 6& 8&10&12&14&16 \\
 counts &15&10&20&20&35&25&36&21&28 \\
 prob &0.0714&0.0476&0.0952&0.0952&0.1667&0.1190&0.1714&0.1000&0.1333 \\
\hline \hline
\end{tabular}
\label{tabRun4b}
%\begin{tablenotes} \item 
\caption*{Distribution conditional on Team A having the first and second runners (No displacement)}
\end{table}
Since this is an American journal, we now evaluate the distribution victory margins under the
IID outcome assumption. The result is in Table \ref{tabRun5} for the case of no displacement.
Conditional on Team A having the fastest runner, the probability of victory is
$72.14\%$ with an average victory of $10.37$. The probability of a loss is $27.86\%$ with a mean loss of $5.837$. There are no ties.
\begin{table} 
\caption{Distribution of score differences Runner:7 Score:5, No Displacement} 
\label{tabRun5}
\centering     %\hline \hline
\begin{tabular}{cccccc  cccccc}
  \hline \hline
$s_B-s_A$ &-15&-13&-11& -9& -7& -5& -3& -1&  1&  3&  5\\
counts & 28& 21& 36& 46& 71& 81& 95&100&125&120&135\\
prob &.0163&.0122&.0210&.0268&.0414&.0472&.0554&.0583&.0728&.0699& .0787\\
\hline
$s_B-s_A$ & 7 &  9& 11& 13& 15& 17& 19& 21& 23& 25 & \\
counts & 125&136&116&117& 92& 95& 64&  49& 28& 36 &\\
prob & .0728&.0793&.0676&.0682&.0536&.0554&.0373&.0286&.0163& .0210 & \\
%\bottomrule[1.25pt]
\hline \hline
\end{tabular}
\caption*{  Distribution conditional on Team A having the first runner. No Displacement.} 
\end{table}
For the symmetric distribution of $|s_A-sB|$, (not conditioning on Team A having the fastest runner), the mean score difference is 
$9.108 \pm 6.283$ with a median value of $9$ and $.9$ quantile of $19$. 
As previously mentioned, the no displacement case is equivalent to the Wilcoxon rank-sum distribution.
The distributions conditional on Team A having the fastest runner appear not to be in the literature.

The distribution for $(7,5)$ with displacement ranges between $-23$ and $35$ conditional on Team $A$ having the fastest runner. Table \ref{tabRun5D} displays this broad distribution. Odd score differences, $s_B-s_A$, are more likely than even differences.
Conditional on Team A having the fastest runner, the probability of victory is
$70.22\%$ with an average victory of $11.66$. The probability of a loss is $28.15\%$ with a mean loss of $6.882$.
\begin{table} %[ht]
  \caption{Distribution of score differences Runner:7 Score:5 with displacement}
  \label{tabRun5D}
\centering     %\hline \hline
\begin{tabular}{ccccccc  cccccc}
  \hline \hline
$s_A -s_B$ &-23&-22&-21&-20&-19&-18&-17&-16&-15&-14&-13&-12 \\
counts & 1& 1& 2& 2& 6& 5& 7& 8&14& 9&18&12 \\
prob &.0006&.0006&.0012&.0012&.0035&.0029&.0041&.0047&.0082&.0052&.0105&.0070 \\
\hline
$s_A -s_B$ &-11&-10& -9& -8& -7& -6& -5& -4& -3& -2& -1&  0 \\
counts &26&16&35&18&47&21&54&23&63&26&69&28 \\
prob &.0152&.0093&.0204&.0105&.0274&.0122&.0315&.0134&.0367&.0152&.0402&.0163 \\
\hline
 $s_A -s_B$ & 1& 2& 3& 4& 5& 6& 7& 8& 9&10&11&12 \\
 counts &79&34&82&31&85&34&83&32&86&33&78&31 \\
 prob &.0460&.0198&.0478&.0181&.0495&.0198&.0484&.0186&.0501&.0192& .0455&.0181 \\
 \hline
 $s_A -s_B$ &13&14&15&16&17&18&19&20&21&22&23&24 \\
 counts &73&32&64&27&55&27&43&24&36&19&26&15 \\
 prob &.0425&.0186&.0373&.0157&.0321&.0157&.0251&.0140&.0210&.0111& .0152&.0087 \\
 \hline
 $s_A -s_B$ &25&26&27&28&29&30&31&32&33&34&35 \\
 counts &21&11&13& 7& 8& 6& 4& 2& 2& 1& 1 \\
 prob &.0122&.0064&.0076&.0041&.0047&.0035&.0023&.0012&.0012&.0006& .0006 \\
%\bottomrule[1.25pt]
\hline \hline
\end{tabular}
\center{Distribution conditional on Team A having the first runner with displacement.} 
\end{table}
%Table \ref{tabRun5D}, just like Table \ref{tabRun5}, gives the distribution assuming that Team A has the fastest runner and with displacement/
Overall (not conditional on Team A having the fastest runner), the mean score difference is 
$10.12 \pm 7.100$ with a median value of $9$. % and $.9$ quantile of $21$.
The $75$th percentile of the score difference is $15$ and the $90$th percentile victory is $21$ for $(7,5)$ .

Similarly, we evaluate the distribution of scoring differences from the IID outcome model when Team A has the fastest two runners. Table \ref {tab75F2D} presents these results for $(7,5)$ with displacement.
Team A wins with a $90.4\%$ probability with an average victory margin of $14.06$ %$13.93$.
Team A loses with a $8.71\%$ probability with average loss of $3.90$. Ties occur with probability $0.9\%$.
\begin{table}
  \caption{Distribution conditional on Team A having the first and second runner $M=7$, $N=5$}
  \label{tab75F2D}
\begin{tabular}{cccccc c cccccc} 
  \hline
  $s_A-s_B$ &-11&-10& -9& -8& -7& -6& -5& -4& -3& -2& -1&  0 \\
 counts & 1& 1& 2& 4& 5& 4&11& 4&12& 7&18& 7 \\
 prob &.0013&.0013&.0025&.0051&.0063&.0051&.0139&.0051&.0152&.0088&
 .0227&.0088 \\
 \hline
 $s_A-s_B$  & 1& 2& 3& 4& 5& 6& 7& 8& 9&10&11&12 \\
 counts &25&13&31&12&34&15&40&15&44&19&45&16 \\
 prob &.0316&.0164&.0391&.0152&.0429&.0189&.0505&.0189&.0556&.0240&
 .0568&.0202 \\
 \hline
 $s_A-s_B$ &13&14&15&16&17&18&19&20&21&22&23&24 \\
 counts &48&20&46&18&41&19&36&19&30&17&24&14 \\
 prob &.0606&.0253&.0581&.0227&.0518&.0240&.0455&.0240&.0379&.0215&
 .0303&.0177 \\
 \hline
 $s_A-s_B$ &25&26&27&28&29&30&31&32&33&34&35 \\
 counts &20&11&13& 7& 8& 6& 4& 2& 2& 1& 1 \\
 prob &.0253&.0139&.0164&.0088&.0101&.0076&.0051&.0025&.0025&.0013&.0013 \\
\hline \hline
\end{tabular}
\caption*{Score difference distribution when Team A has the fastest two runners with displacement, $M=7$, $N=5$.}
\end{table}
If one team has the fastest runner and the other team has the second fastest runner, the distribution
of scores for $(M,N)$ then coresponds to a shifted version of $(M,N)$. In particular for $(7,5)$, let Team A
be the team with the third fastest runner. The distribution of victory margins is given by Table \ref{tabRun4}
shifted by one. If Team A has the first and third fastest runners, add one to the score differences in Table \ref{tabRun4}.
If Team A has the second and third fastest runners, subtract one to the score differences in Table \ref{tabRun4}.
%Since the score differences for Table \ref{tabRun4}

\section{Population Randomness Only} %{POPULATION RANDOMNESS ONLY}

Our first model postulates that all runners are equally likely to win or come in last. This models the situation
where there is a large amount of variation in each runner's time relative to the skill difference between
runners. If, on the other hand, the standing of the runners did not vary and the results were predetermined
before the start of the race, what would matter is the size of the population from which the runners are
chosen.

Assume that School A has a population of $M_A$ potential runners and School B has a population of $M_B$ potential runners.
The speeds of the potential runners are randomly distributed over the population of $M_A + M_B$ potential runners.
We assume that each coach can perfectly identify the best $M$ runners.
%(An assumption that is not always satisfied in the Putnam competition.)

In this random population model, the probability that Team A has the fastest runner (or any other runner) is $M_A/(M_A +M_B)$.
One proxy for $M_A$ is the number of students at a school times some constant that corresponds to the participation rate.
The participation rate is larger for smaller schools, but we can assume that good coaches always recruit
the fastest runners and the recruiting population is proportional to school size. Each coach then chooses the best $M$
runners to compete.

This random population model is parameterized by the tuple $(M_A,M_B,M,N)$. The random outcome model is a special case with
$M_A = M_B = M$. Of course, in the random population model, the rankings are predetermined before the race starts by the abilities
of each runner. For each value of $(M_A,M_B,M,N)$, we could calculate the distribution of results. Instead we consider the limit
of $M_A >> M$ and $M_B >>M$ such that the ratio $M_A/(M_A +M_B)$ tends to a constant fraction, $r$. In this limit, the score distribution
converges to the distribution which corresponds to drawing runners with replacement. 

One very convenient property of cross-country scoring is that the final score depends only on the results of the runners
who come in in the first $2N -1$ places. Thus we need to compute probabilities of the $2^{2N-1}$ racing scenarios.
As $M_A$ and $M_B$ increase, the probability of any of the scenarios converges to $r^k (1-r)^{2N-k-1}$ where $k$ is the number of runners from Team A.

%The case of drawing with replacement corresponds to two very large teams of equal size.
%The more general formulation corresponding to poulation randomness only is Team A has $M_A$ potential runners
%and Team B has $M_B$ potential runners and these are ranked. Each team then drops all but its top $N$ runners
%and these $2 N$ runners are ranked.
%Let us assume that $N_A$ and $N_B$ are much larger than $K$, the number of runners which count in the scoring.
%Let $P = N_A/(N_A +N_B)$ and the probability of Team A getting the $j$th runner before dropping runners is $P$.

In the large population limit of the random population model with no displacement, the results depend only on $N$ and $r$. We tabulate the score difference distribution as a function of $r =M_A/(M_A +M_B)$ for $N=4$ and $N=5$. Once again, we consider both the case of displacement and the simpler case of no displacement. Tables \ref{Stat4}-\ref{Stat5D} present the summary statistics  for the large population limit.
In Tables \ref{Stat4}- \ref{Stat5D}, {\it Prob of Win} is the probability that Team A wins,
{\it  Mn score Diff} is the mean score difference i.e. the expectation of $s_A -s_B$,
{\it Std Dev Diff} is the standard deviation of $s_A -s_B$, {\it Mn Win Diff} is
the expectation of $s_A -s_B$ conditional on Team A winning and  {\it Mn Loss Diff} is
the expectation of $s_A -s_B$ conditional on Team A losing. The final row is our large victory margin, the $quantile(.9)$.
For this, we are computing the quantile for the absolute victory margin. Ten percent of the time the there will be a victory margin of this size, but it may be a loss for Team A.

%such as mean score difference and its standard deviation, probability of a win,
%mean victory margin conditional on a win, mean loss margin conditional on a loss amd the score difference at quantile equal $.9$.
\begin{table}[ht]
\caption{Statistics of Score Difference: Scorers: $4$, No Displacement}
% title of Table
\centering 
\begin{tabular}{c c c c  c c c c }
% centered columns (4 columns)
%\[0.5ex]
\hline \hline
Ratio $r$ &  0.5 & 0.55& 0.6 & 0.65& 0.7 & 0.75& 0.8 \\
\hline
 Prob of Win &.4609&.5541&.6452&.7306&.8070&.8718&.9232 \\
 Mn score Diff &0.00&2.03&4.02&5.94&7.76&9.47&11.05 \\
 Std Dev Diff &9.24&9.14&8.83&8.33&7.67&6.86&5.94 \\
 Mn Diff Win&8.47&8.99&9.53&10.13&10.78&11.49&12.26 \\
 Mn Diff Loss&-8.47&-7.99&-7.54&-7.10&-6.67&-6.25&-5.83 \\
 Quantile(.9) &12&14&16&16&16&16&16 \\
 \hline
\end{tabular}
\label{Stat4}
\caption*{Victory margin for four scorers with no displacement.
  Ratio is  the fraction of total population for Team A.}
%\begin{tablenotes}\item \end{tablenotes}
\end{table}
Both the mean victory margin and the standard deviation are larger for the case of displacement as seen by comparing Tables \ref{Stat4}
and \ref{Stat4D}.
\begin{table}[ht]
\caption{Statistics of Score Difference: $M=6$, $N=4$, Displacement.}
% title of Table
\centering 
\begin{tabular}{c c c c  c c c c }
% centered columns (4 columns)
%\[0.5ex]
\hline \hline
Ratio $r$ &  0.5 & 0.55& 0.6 & 0.65& 0.7 & 0.75& 0.8 \\
\hline
 Prob. of  Win &0.4785&0.5785&0.6749&0.7632&0.8395&0.9011&0.9467 \\
 Mean Score Diff &0.00&2.94&5.85&8.68&11.39&13.96&16.35 \\
 Std Dev Diff&12.04&11.92&11.56&10.97&10.16&9.16&7.99 \\
 Mean Win Diff &10.51&11.44&12.45&13.56&14.78&16.11&17.55 \\
 Mean Loss Diff &-10.51&-9.67&-8.89&-8.18&-7.51&-6.89&-6.29 \\
 Quantile(.9) &16&18&21&23&24&24&24 \\
 \hline
\end{tabular}
\label{Stat4D}
\caption*{
  Victory margin for six runners, four scorers with displacement. Ratio is the fraction of total population 
  for Team A.}
%\begin{tablenotes}\item \end{tablenotes}
\end{table}

\begin{table}[ht] 
\caption{Statistics of Score Difference: N=5, no Displacement} \label{Stat5}
\centering 
\begin{tabular}{c c c c  c c c c }
%\ [0.5ex]
\hline \hline
Ratio $r$ &  0.5 & 0.55& 0.6 & 0.65& 0.7 & 0.75& 0.8 \\
\hline
 Prob of Win &.5000&.6037&.7018&.7892&.8623&.9187&.9584 \\
 Mean Score Diff &0.00&3.28&6.49&9.56&12.45&15.11&17.53 \\
 Std Dev Diff  &13.28&13.10&12.60&11.80&10.75&9.52&8.16 \\
 Mean Win Diff&11.21&12.19&13.26&14.42&15.68&17.07&18.56 \\
 Mean Loss Diff&-11.21&-10.30&-9.43&-8.60&-7.80&-7.00&-6.19 \\
 Quantile(.9)&17 &21&23&25 &25 & 25 &25 \\
%  Quantile(.9) & 17 &21&23&25 &25 & 25 &25 \\ 
\hline
\end{tabular}
\caption*{ Victory margin for five scorers score versus the fraction of total population.} %for Team A. }
\end{table}

\begin{table}[ht] 
\caption{Statistics of Score Difference: $M=7$, $N=5$, Displacement} \label{Stat5D}
\centering 
\begin{tabular}{c c c c  c c c c }
\hline \hline
Ratio $r$ &  0.5 & 0.55& 0.6 & 0.65& 0.7 & 0.75& 0.8 \\
\hline
 Prob. of  Win &0.4951&0.6052&0.7089&0.8002&0.8746&0.9301&0.9670 \\
 Mean Score Diff &0.00&4.48&8.88&13.12&17.14&20.88&24.31 \\
 Std Dev  &16.58&16.39&15.81&14.88&13.64&12.14&10.47 \\
 Mean Win Diff &13.96&15.44&17.07&18.88&20.87&23.04&25.38 \\
 Mean Loss Diff &-13.96&-12.62&-11.39&-10.26&-9.21&-8.20&-7.21 \\
 Quantile(.9) &23&26&29&31&34&35&35 \\
\hline
\end{tabular}
\caption*{ Victory margin for seven runners, five scorers with displacement. Ratio is the fraction of total population for Team A. }
\end{table}

For completeness, we give the score difference distribution as a function of $r$ in Tables \ref{Prob4} and \ref{Prob5}. Note that $s_A -s_B=0$ is the probability of a tie in Table \ref{Prob4}.
\begin{table}[!ht]
\caption{Distribution of Score Difference: $N=4$, No Displacement}
% title of Table
\centering 
\begin{tabular}{cccc cccc}
\hline \hline
$s_B-s_A$  &  0.5 & 0.55& 0.6 & 0.65& 0.7 & 0.75& 0.8 \\
\hline
-16 &0.0625&0.0410&0.0256&0.0150&0.0081&0.0039&0.0016 \\
 -14 &0.0312&0.0226&0.0154&0.0098&0.0057&0.0029&0.0013 \\
 -12 &0.0469&0.0350&0.0246&0.0161&0.0096&0.0051&0.0023 \\
 -10 &0.0547&0.0418&0.0301&0.0202&0.0124&0.0068&0.0031 \\
 -8 &0.0781&0.0625&0.0476&0.0342&0.0229&0.0139&0.0074 \\
 -6 &0.0547&0.0468&0.0378&0.0286&0.0200&0.0126&0.0070 \\
 -4 &0.0703&0.0620&0.0516&0.0404&0.0292&0.0192&0.0111 \\
 -2 &0.0625&0.0579&0.0507&0.0417&0.0318&0.0220&0.0133 \\
 0 &0.0781&0.0764&0.0714&0.0635&0.0534&0.0417&0.0297 \\
 2 &0.0625&0.0640&0.0622&0.0572&0.0494&0.0396&0.0287 \\
 4 &0.0703&0.0757&0.0774&0.0749&0.0682&0.0577&0.0442 \\
 6 &0.0547&0.0606&0.0636&0.0632&0.0590&0.0511&0.0401 \\
 8 &0.0781&0.0934&0.1071&0.1180&0.1245&0.1252&0.1188 \\
 10 &0.0547&0.0680&0.0809&0.0920&0.1001&0.1038&0.1016 \\
 12 &0.0469&0.0597&0.0726&0.0843&0.0936&0.0989&0.0983 \\
 14 &0.0312&0.0412&0.0518&0.0625&0.0720&0.0791&0.0819 \\
 16 &0.0625&0.0915&0.1296&0.1785&0.2401&0.3164&0.4096 \\
\hline
\end{tabular}
\label{Prob4}
\caption*{Victory margin for four runners count versus the fraction of total population for Team A. No Displacement.}
\end{table}
For international dual-meets, Table \ref{Prob4} shows that more often than not the victory margin  is divisible by four.
Specifically, for $(6,4)$, the probability that the victory margin is divisible by four is .5625. For the random population problem with ratio $r=.5$, the probability that the victory margin is divisble by four is $.59375$.

\begin{table}[!ht]
\caption{Distribution of Score Difference:$M=6$, $N=4$ Displacement}
% title of Table
\centering 
\begin{tabular}{c c c c  c c c c }
%\ [0.5ex]
\hline \hline
$s_B-s_A$  &  0.5 & 0.55& 0.6 & 0.65& 0.7 & 0.75& 0.8 \\
-24 &0.0156&0.0083&0.0041&0.0018&0.0007&0.0002&0.0001 \\
 -23 &0.0078&0.0046&0.0025&0.0012&0.0005&0.0002&0.0001 \\
 -22 &0.0117&0.0071&0.0039&0.0020&0.0009&0.0003&0.0001 \\
 -21 &0.0059&0.0039&0.0024&0.0013&0.0006&0.0002&0.0001 \\
 -20 &0.0156&0.0101&0.0061&0.0034&0.0017&0.0007&0.0003 \\
 -19 &0.0078&0.0056&0.0037&0.0022&0.0012&0.0005&0.0002 \\
 -18 &0.0176&0.0115&0.0070&0.0039&0.0020&0.0008&0.0003 \\
 -17 &0.0098&0.0073&0.0050&0.0032&0.0018&0.0009&0.0003 \\
 -16 &0.0254&0.0178&0.0118&0.0074&0.0043&0.0022&0.0010 \\
 -15 &0.0117&0.0087&0.0059&0.0037&0.0020&0.0010&0.0004 \\
 -14 &0.0293&0.0209&0.0140&0.0088&0.0051&0.0026&0.0011 \\
 -13 &0.0137&0.0103&0.0072&0.0046&0.0026&0.0013&0.0005 \\
 -12 &0.0312&0.0239&0.0173&0.0116&0.0071&0.0039&0.0018 \\
 -11 &0.0137&0.0109&0.0080&0.0053&0.0031&0.0015&0.0006 \\
 -10 &0.0332&0.0269&0.0204&0.0144&0.0093&0.0053&0.0026 \\
 -9 &0.0117&0.0095&0.0071&0.0048&0.0028&0.0014&0.0006 \\
 -8 &0.0391&0.0327&0.0257&0.0188&0.0126&0.0075&0.0038 \\
 -7 &0.0137&0.0116&0.0091&0.0065&0.0042&0.0024&0.0011 \\
 -6 &0.0391&0.0341&0.0281&0.0218&0.0157&0.0102&0.0058 \\
 -5 &0.0137&0.0119&0.0095&0.0069&0.0045&0.0026&0.0012 \\
 -4 &0.0430&0.0385&0.0326&0.0257&0.0187&0.0123&0.0070 \\
 -3 &0.0137&0.0123&0.0102&0.0077&0.0053&0.0032&0.0016 \\
 -2 &0.0410&0.0385&0.0341&0.0282&0.0216&0.0149&0.0090 \\
 -1 &0.0137&0.0129&0.0113&0.0090&0.0064&0.0040&0.0021 \\
\hline
\end{tabular}
\label{Prob4D}
\caption*{Victory margin for six runners, four scorers versus the fraction of total population for Team A with displacement. (Part 1)}
\end{table}

\begin{table}[!ht]
\caption{Distribution of Score Difference:$M=6$, $N=4$ Displacement}
% title of Table
\centering 
\begin{tabular}{c c c c  c c c c }
%\ [0.5ex]
\hline \hline
$s_B-s_A$  &  0.5 & 0.55& 0.6 & 0.65& 0.7 & 0.75& 0.8 \\ 
 0 &0.0430&0.0417&0.0382&0.0326&0.0258&0.0185&0.0116 \\
 1 &0.0137&0.0133&0.0119&0.0098&0.0072&0.0046&0.0025 \\
 2 &0.0410&0.0412&0.0389&0.0344&0.0282&0.0209&0.0136 \\
 3 &0.0137&0.0142&0.0136&0.0121&0.0098&0.0071&0.0045 \\
 4 &0.0430&0.0454&0.0455&0.0432&0.0385&0.0319&0.0240 \\
 5 &0.0137&0.0145&0.0143&0.0129&0.0106&0.0077&0.0048 \\
 6 &0.0391&0.0425&0.0438&0.0428&0.0392&0.0332&0.0253 \\
 7 &0.0137&0.0150&0.0153&0.0144&0.0124&0.0096&0.0064 \\
 8 &0.0391&0.0442&0.0474&0.0483&0.0463&0.0414&0.0337 \\
 9 &0.0117&0.0133&0.0139&0.0134&0.0118&0.0093&0.0063 \\
 10 &0.0332&0.0387&0.0428&0.0447&0.0438&0.0399&0.0329 \\
 11 &0.0137&0.0158&0.0169&0.0167&0.0150&0.0121&0.0084 \\
 12 &0.0312&0.0385&0.0450&0.0495&0.0511&0.0488&0.0422 \\
 13 &0.0137&0.0168&0.0194&0.0209&0.0210&0.0195&0.0163 \\
 14 &0.0293&0.0391&0.0498&0.0607&0.0707&0.0782&0.0814 \\
 15 &0.0117&0.0148&0.0174&0.0192&0.0197&0.0185&0.0157 \\
 16 &0.0254&0.0345&0.0449&0.0558&0.0661&0.0745&0.0788 \\
 17 &0.0098&0.0123&0.0144&0.0160&0.0165&0.0158&0.0136 \\
 18 &0.0176&0.0252&0.0341&0.0438&0.0536&0.0621&0.0676 \\
 19 &0.0078&0.0102&0.0124&0.0142&0.0151&0.0148&0.0131 \\
 20 &0.0156&0.0226&0.0311&0.0406&0.0504&0.0593&0.0655 \\
 21 &0.0059&0.0081&0.0105&0.0125&0.0138&0.0139&0.0126 \\
 22 &0.0117&0.0181&0.0261&0.0356&0.0459&0.0556&0.0629 \\
 23 &0.0078&0.0125&0.0187&0.0264&0.0353&0.0445&0.0524 \\
 24 &0.0156&0.0277&0.0467&0.0754&0.1176&0.1780&0.2621 \\
\hline
\end{tabular}
\label{Prob4Db}
\caption*{Victory margin for six runners, four scorers versus the fraction of total population for Team A with displacement. Continued. (Part 2)}
\end{table}

\begin{table} %[ht]
\caption{Distribution of Score Differences: $N=5$ No Displacement}
% title of Table
\centering 
\begin{tabular}{c c c c  c c c c }
% centered columns (8 columns)
%\ [0.5ex]
  \hline \hline
  $s_B-s_A$  &  0.5 & 0.55& 0.6 & 0.65& 0.7 & 0.75& 0.8 \\
\hline
-25 &0.0312&0.0185&0.0102&0.0053&0.0024&0.0010&0.0003 \\
 -23 &0.0156&0.0101&0.0061&0.0034&0.0017&0.0007&0.0003 \\
 -21 &0.0234&0.0157&0.0098&0.0056&0.0029&0.0013&0.0005 \\
 -19 &0.0273&0.0188&0.0120&0.0071&0.0037&0.0017&0.0006 \\
 -17 &0.0371&0.0261&0.0171&0.0102&0.0055&0.0026&0.0010 \\
 -15 &0.0430&0.0312&0.0213&0.0134&0.0077&0.0039&0.0016 \\
 -13 &0.0410&0.0314&0.0223&0.0146&0.0086&0.0044&0.0019 \\
 -11 &0.0410&0.0326&0.0242&0.0165&0.0102&0.0055&0.0025 \\
 -9 &0.0469&0.0381&0.0288&0.0201&0.0127&0.0071&0.0033 \\
 -7 &0.0469&0.0397&0.0313&0.0228&0.0151&0.0088&0.0043 \\
 -5 &0.0508&0.0446&0.0368&0.0282&0.0200&0.0127&0.0070 \\
 -3 &0.0469&0.0431&0.0370&0.0295&0.0215&0.0140&0.0079 \\
 -1 &0.0488&0.0463&0.0411&0.0340&0.0258&0.0176&0.0105 \\
 1 &0.0488&0.0482&0.0446&0.0383&0.0303&0.0215&0.0133 \\
 3 &0.0469&0.0477&0.0453&0.0401&0.0326&0.0239&0.0152 \\
 5 &0.0508&0.0545&0.0551&0.0524&0.0466&0.0381&0.0281 \\
 7 &0.0469&0.0519&0.0539&0.0525&0.0475&0.0396&0.0295 \\
 9 &0.0469&0.0541&0.0587&0.0598&0.0570&0.0501&0.0397 \\
 11 &0.0410&0.0483&0.0534&0.0554&0.0537&0.0479&0.0385 \\
 13 &0.0410&0.0501&0.0576&0.0620&0.0623&0.0578&0.0484 \\
 15 &0.0430&0.0560&0.0693&0.0817&0.0917&0.0976&0.0976 \\
 17 &0.0371&0.0497&0.0630&0.0758&0.0866&0.0936&0.0949 \\
 19 &0.0273&0.0374&0.0485&0.0598&0.0701&0.0779&0.0813 \\
 21 &0.0234&0.0328&0.0435&0.0548&0.0655&0.0742&0.0786 \\
 23 &0.0156&0.0226&0.0311&0.0406&0.0504&0.0593&0.0655 \\
 25 &0.0312&0.0503&0.0778&0.1160&0.1681&0.2373&0.3277 \\
\hline
\end{tabular}
\label{Prob5}
\caption*{Victory margin for five scorers with no displacement versus the fraction of total population in Team A.}
\end{table}

\begin{table} %[ht]
\caption{Distribution of Score Differences:$M=7$, $N=5$ with displacement}
% title of Table
\centering 
\begin{tabular}{c c c c  c c c c }
% centered columns (8 columns)
%\ [0.5ex]
  \hline \hline
  $s_B-s_A$  &  0.5 & 0.55& 0.6 & 0.65& 0.7 & 0.75& 0.8 \\
  \hline
-35 &0.0078&0.0037&0.0016&0.0006&0.0002&0.0001&0  \\
 -34 &0.0039&0.0021&0.0010&0.0004&0.0002&0 &0  \\
 -33 &0.0059&0.0032&0.0016&0.0007&0.0003&0.0001&0  \\
 -32 &0.0029&0.0018&0.0009&0.0004&0.0002&0.0001&0  \\
 -31 &0.0073&0.0041&0.0021&0.0010&0.0004&0.0001&0  \\
 -30 &0.0039&0.0025&0.0015&0.0008&0.0004&0.0001&0  \\
 -29 &0.0088&0.0052&0.0028&0.0014&0.0006&0.0002&0.0001 \\
 -28 &0.0044&0.0029&0.0017&0.0009&0.0004&0.0002&0  \\
 -27 &0.0117&0.0071&0.0039&0.0020&0.0009&0.0003&0.0001 \\
 -26 &0.0059&0.0039&0.0024&0.0013&0.0006&0.0002&0.0001 \\
 -25 &0.0142&0.0090&0.0053&0.0029&0.0014&0.0006&0.0002 \\
 -24 &0.0068&0.0047&0.0029&0.0016&0.0008&0.0003&0.0001 \\
 -23 &0.0181&0.0115&0.0068&0.0036&0.0018&0.0007&0.0002 \\
 -22 &0.0088&0.0060&0.0038&0.0021&0.0010&0.0004&0.0001 \\
 -21 &0.0176&0.0121&0.0077&0.0045&0.0023&0.0010&0.0004 \\
 -20 &0.0083&0.0061&0.0040&0.0024&0.0012&0.0005&0.0002 \\
 -19 &0.0200&0.0141&0.0092&0.0055&0.0030&0.0014&0.0005 \\
 -18 &0.0093&0.0069&0.0047&0.0028&0.0015&0.0006&0.0002 \\
 -17 &0.0215&0.0158&0.0108&0.0068&0.0038&0.0018&0.0007 \\
 -16 &0.0093&0.0071&0.0048&0.0030&0.0016&0.0007&0.0002 \\
 -15 &0.0244&0.0184&0.0129&0.0083&0.0048&0.0024&0.0010 \\
 -14 &0.0103&0.0080&0.0056&0.0036&0.0020&0.0009&0.0003 \\
 -13 &0.0254&0.0201&0.0148&0.0101&0.0062&0.0034&0.0015 \\
 -12 &0.0098&0.0078&0.0056&0.0036&0.0020&0.0010&0.0004 \\
 -11 &0.0278&0.0223&0.0166&0.0114&0.0071&0.0038&0.0017 \\
 -10 &0.0107&0.0087&0.0064&0.0042&0.0024&0.0012&0.0005 \\
 -9 &0.0283&0.0240&0.0188&0.0135&0.0088&0.0050&0.0024 \\
 -8 &0.0098&0.0083&0.0063&0.0043&0.0026&0.0013&0.0005 \\
 -7 &0.0293&0.0254&0.0204&0.0151&0.0101&0.0059&0.0029 \\
 -6 &0.0103&0.0089&0.0071&0.0050&0.0031&0.0016&0.0007 \\
 -5 &0.0303&0.0271&0.0226&0.0173&0.0120&0.0073&0.0037 \\
 -4 &0.0098&0.0087&0.0070&0.0051&0.0032&0.0017&0.0007 \\
 -3 &0.0308&0.0282&0.0239&0.0186&0.0131&0.0081&0.0042 \\
 -2 &0.0107&0.0100&0.0085&0.0065&0.0044&0.0026&0.0012 \\
 -1 &0.0312&0.0300&0.0268&0.0222&0.0168&0.0114&0.0066 \\
\hline
\end{tabular}
\label{Prob5D}
\caption*{Victory margin for seven runners, five scorers with displacement versus the fraction of total population. (Part 1)}
\end{table}

 \begin{table} %[ht]
\caption{Distribution of Score Differences:$M=7$, $N=5$ with displacement}
\centering 
\begin{tabular}{c c c c  c c c c }
% centered columns (8 columns)
%\ [0.5ex]
  \hline \hline
  $s_B-s_A$  &  0.5 & 0.55& 0.6 & 0.65& 0.7 & 0.75& 0.8 \\
\hline 
 0 &0.0098&0.0093&0.0081&0.0063&0.0044&0.0026&0.0013 \\
 1 &0.0312&0.0304&0.0275&0.0230&0.0176&0.0120&0.0070 \\
 2 &0.0107&0.0106&0.0094&0.0076&0.0055&0.0034&0.0018 \\
 3 &0.0308&0.0313&0.0296&0.0260&0.0208&0.0150&0.0093 \\
 4 &0.0098&0.0100&0.0093&0.0078&0.0059&0.0038&0.0021 \\
 5 &0.0303&0.0315&0.0306&0.0275&0.0228&0.0170&0.0110 \\
 6 &0.0103&0.0107&0.0102&0.0088&0.0067&0.0045&0.0025 \\
 7 &0.0293&0.0315&0.0315&0.0293&0.0250&0.0192&0.0129 \\
 8 &0.0098&0.0105&0.0103&0.0091&0.0071&0.0049&0.0027 \\
 9 &0.0283&0.0312&0.0320&0.0305&0.0267&0.0210&0.0144 \\
 10 &0.0107&0.0121&0.0125&0.0117&0.0099&0.0074&0.0047 \\
 11 &0.0278&0.0325&0.0357&0.0366&0.0350&0.0307&0.0241 \\
 12 &0.0098&0.0112&0.0117&0.0111&0.0095&0.0072&0.0046 \\
 13 &0.0254&0.0300&0.0334&0.0347&0.0335&0.0297&0.0236 \\
 14 &0.0103&0.0120&0.0130&0.0128&0.0114&0.0091&0.0062 \\
 15 &0.0244&0.0303&0.0353&0.0386&0.0394&0.0371&0.0314 \\
 16 &0.0093&0.0111&0.0122&0.0122&0.0110&0.0089&0.0061 \\
 17 &0.0215&0.0273&0.0325&0.0363&0.0376&0.0359&0.0308 \\
 18 &0.0093&0.0114&0.0128&0.0132&0.0123&0.0103&0.0074 \\
 19 &0.0200&0.0265&0.0329&0.0382&0.0413&0.0412&0.0370 \\
 20 &0.0083&0.0103&0.0118&0.0124&0.0117&0.0099&0.0072 \\
 21 &0.0176&0.0238&0.0302&0.0357&0.0393&0.0397&0.0361 \\
 22 &0.0088&0.0118&0.0146&0.0168&0.0179&0.0174&0.0151 \\
 23 &0.0181&0.0266&0.0369&0.0483&0.0598&0.0696&0.0755 \\
 24 &0.0068&0.0093&0.0116&0.0136&0.0147&0.0146&0.0130 \\
 25 &0.0142&0.0210&0.0294&0.0391&0.0492&0.0585&0.0650 \\
 26 &0.0059&0.0081&0.0105&0.0125&0.0138&0.0139&0.0126 \\
 27 &0.0117&0.0181&0.0261&0.0356&0.0459&0.0556&0.0629 \\
 28 &0.0044&0.0062&0.0082&0.0100&0.0113&0.0116&0.0108 \\
 29 &0.0088&0.0138&0.0205&0.0285&0.0375&0.0466&0.0541 \\
 30 &0.0039&0.0056&0.0075&0.0092&0.0106&0.0111&0.0105 \\
 31 &0.0073&0.0119&0.0182&0.0260&0.0350&0.0443&0.0523 \\
 32 &0.0029&0.0045&0.0063&0.0081&0.0096&0.0104&0.0101 \\
 33 &0.0059&0.0099&0.0157&0.0232&0.0321&0.0417&0.0503 \\
 34 &0.0039&0.0069&0.0112&0.0172&0.0247&0.0334&0.0419 \\
 35 &0.0078&0.0152&0.0280&0.0490&0.0824&0.1335&0.2097 \\
\hline
\end{tabular}
\label{Prob5Db}
\caption*{Victory margin for five scorers, seven runners with displacement versus the fraction of total population. Continued.}
\end{table}

\begin{figure}
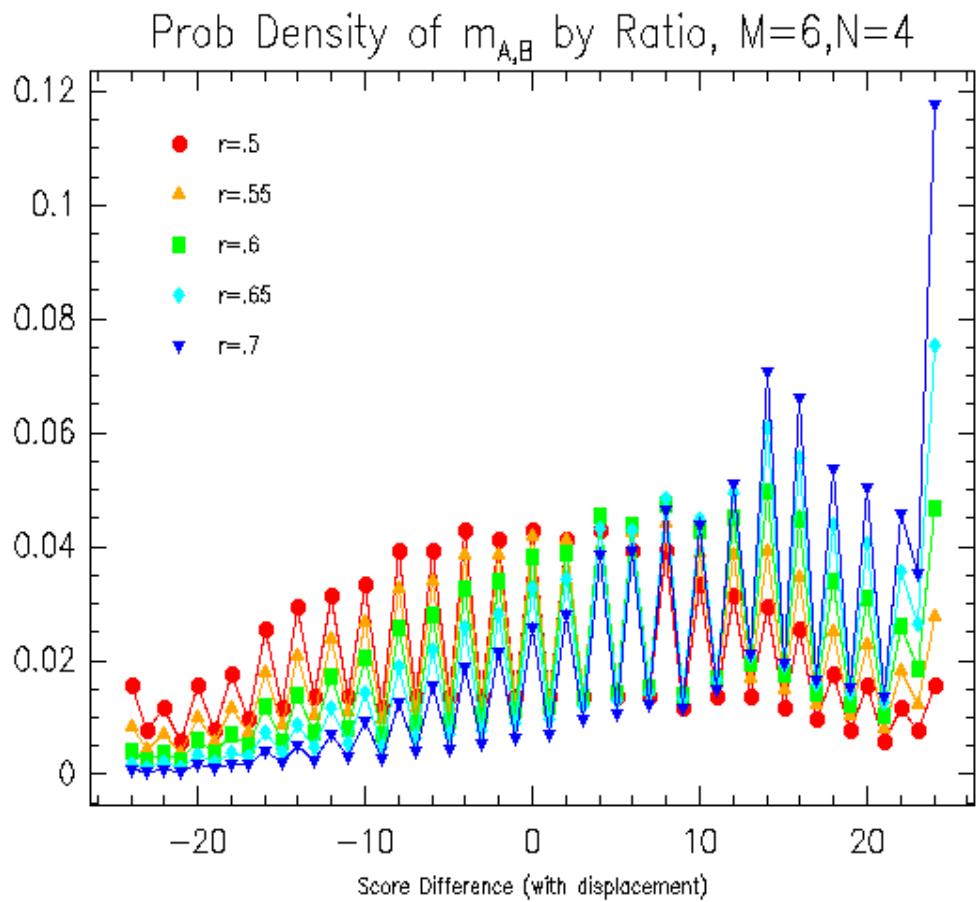
 %[h]
  \centering  
  \caption{Probability density of score difference for four scorers, six runners with displacement.} \label{Fig_R4D}    
\end{figure}

\begin{figure}
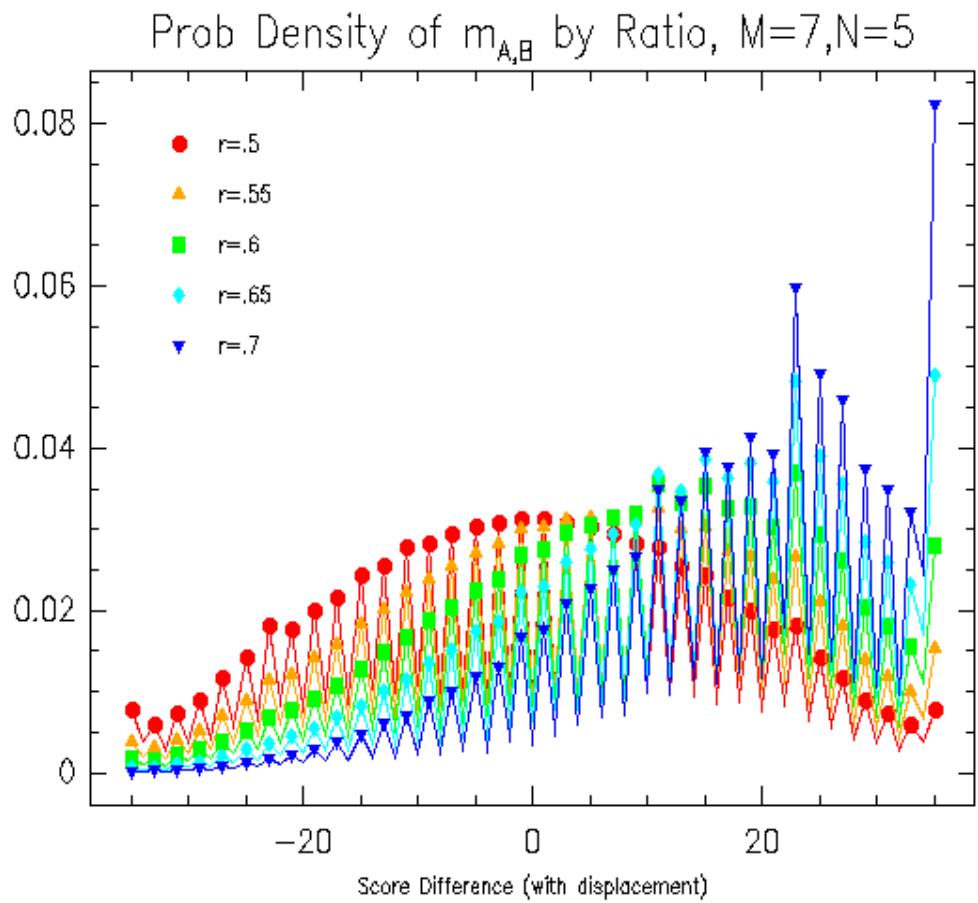
 %[h]
  \centering
  \caption{Probability density of score difference for five scorers, seven runners with displacement.}
  \label{Fig_R5D}    
\end{figure}

 Since a picture is worth a thousand words, Figure \ref{Fig_R4}-\ref{Fig_R5D}  plot the distributions of  Table \ref{Prob4}-\ref{Prob5D}. For $N=4$, $r=.55$, $m_{A,B}=8$ is the most probable value. As $r$ increases, the primary maximum shifts to the largest possible value, $16$, with a secondary maximum at 8.
\begin{figure} %[h]
  \centering 
  \caption{Probability density of score difference for four scorers, no displacement.}  \label{Fig_R4}  
\end{figure}
Figure \ref{Fig_R4} appears smooth while Figure \ref{Fig_R4D} appears highly oscillatory. This is an illusion as the odd numbered values are identically zero for the no displacement case. These same values are small but nonzero for the
case with displacement. Our input into the plotting package for Figure \ref{Fig_R4} does not give the zero values at the odd integers.
\begin{figure}
  \centering  
  \caption{Probability density of score difference for five scorers, no displacement}
  \label{Fig_R5}  
\end{figure}
Figure \ref{Fig_R5} appears smooth because the even numbered values are identically zero for the no displacement case with $M=5$. For $M=5$, the secondary relative maximum of the density occurs at $m_{A,B}=15$.

\section{Other Models and Extensions} %{OTHER MODELS AND EXTENSION}

We now discuss possible generalizations and alternative models. We can model the effect of an injury during the
race such as a pulled hamstring by calculating the score distribution of $(M,M-1,N)$ such as $(7,6,5)$.

The models of Sections II and III are simple to explain and can be computed easily numerically. To get models that can be easily evaluated
analytically, we can make more restrictive assumptions on the distribution of speeds. A very simple assumption is that
the top runners on both teams are of equal ability and much faster than all other runners. We make this assumption iteratively
on the remaining runners on the team. In this scenario, the distribution of scores corresponds to $N$ independent races.
The score differences are then a shifted scaling binomial distribution. (The shifting and scaling occur because the score differences are $\{-1, 1\}$,
not $\{0,1\}$.) Another tractable model is to assume that the top two runners of each team are much faster than all other runners
and next two runners of each team are much faster than the rest of the runners. This essentially decomposes the scores for
a $(M,N)$ race into two independent $(2,2)$ races plus the residual scoring from the remainder of the team.
These last two examples illustrate the ease with which one can create new scenarios for cross-country models

Given actual race time data for each runner, we can create probabilistic models for distribution of race times for each runner.
We continue to assume  that the runners' times are independent. Of course, this ignores the competitive nature of racing: If
another runner is running fast, you are likely to run faster even at the risk of burning yourself out and possibly finishing much worse.
We could fit each runner's race times to a density, $p_i(t; c_i)$ where the free parameters $c_i$ are fit to the data.
The probability that runner $i$ is faster than runner $j$ is given by
$Prob(T_i < T_j) = \int P_i(t; c_i) p_j(t; c_j) \mathrm{d}t $ where $P_i$ is the integral of $p_i$: 
$P_i(t; c_i) \equiv \int_0^t  p_i(s; c_i) ds$.
The simplest model is that the $i$th runner's time is uniformly distributed on the interval,
$[b_i,B_i]$. Given the values of $b_i$ and $B_i$ for each runner, one can compute the probability of each possible ordering
either evaluating the integrals explicity (I assume using a symbolic manipulation program) or using Monte Carlo.
I believe that the running times are probably more peaked about the mode with a long tail corresponding to ``off'' days.
Thus a shifted stretched beta distribution with $a \approx 1.5$ and $b \approx 3$ may be a more realistic model. The shifted stretched
beta distribution has four free parameters and this is often too many.

Not all races or race courses are equally easy. Hilly race courses and unpleasant race conditions will slow down all runners.
Any attempt to fit empirical data will likely benefit from standardizing the times by race or at least by race course.
I have personal experience with a runner who only wins on hilly race courses. Of course, modelling this would again add an
extra free parameter to each density, $p_i(t; c_i)$, and is likely not worthwhile.

\section{Summary}

\begin{figure}
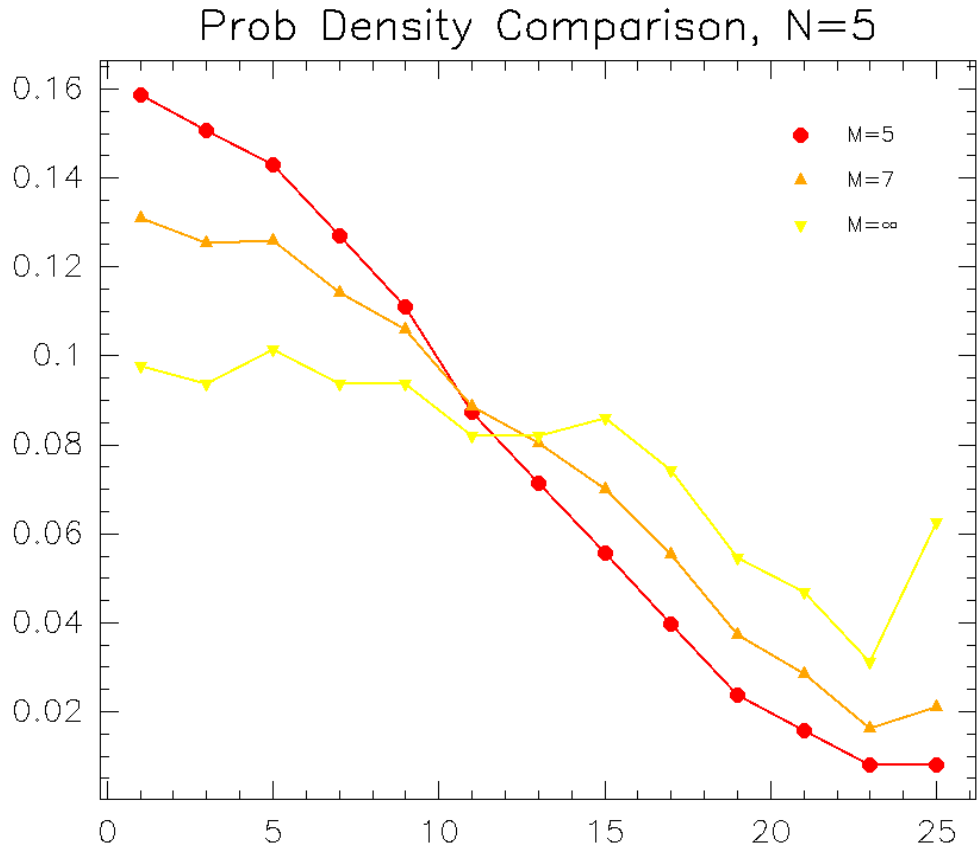
 %[h]
  \centering
  %\caption{} %Figure 2: $N=5$}
  %\includegraphics[width=.8\columnwidth,natwidth=9cm]{{CCFig4}.pdf}    
\caption{Score difference, $|s_A-s_B|$, for the IID Outcome with $N=5$ and no displacement versus the number of runners, $M$.}
  \label{Fig_C5}
\end{figure}

The two models presented above, the IID Outcome model and the Large Population model
are baseline models models from which we get interesting and very testable score distributions.
The population model gives a probability of victory when two schools of different sizes compete.
Given a database of dual-meet scores and school sizes, it would be of interest to see how well the
predictions in Tables 5 - 8 %\ref{Stat4,Stat5}
are supported by the data. Our tabulated distributions show that as $M$ increases at fixed $N$, the probability of a large victory increases. Figure \ref{Fig_C5} shows the distributions for $N=5$ as M increases from $M=5$
to $M=7$ to $M=\infty$ for the case of no no displacement..
We defined a big victory to be a victory in the 90{\it th} percentile of victories. 
For the IID Outcome model, a big victory is $|m_{A,B}|=15$ with displacement and $|m_{A,B}|=14$ without for $(6,4)$ (International dual-meets).
For American dual-meets, $(7,5)$, the $.9$ quantile is $|m_{A,B}|=21$ with displacement
and $|m_{A,B}|=19$ without.

We have analyzed the statistical properties of rank sum scoring with and without scoring. The true argument in favor
of scoring with displacement is to motivate the slower runners so that they can favorably impact the success of the team.

%%http://www.mathaware.org/mam/2010/essays/SzydlikCrossCountry.pdf
\section{Acknowledgement}
We thank the anonymmous referee for pointing out the Wilcoxon rank-sum statistic as well as many other references.

\end{document}